\documentclass[aps,prd,reprint,superscriptaddress]{revtex4-2}
\usepackage{amsfonts,amsmath,amssymb}
\usepackage{bm}
\usepackage{booktabs} 
\usepackage{graphicx}
\usepackage{xcolor}
\usepackage[percent]{overpic}
\usepackage{mathtools,upgreek,pifont}

\usepackage{accents}

\begin{document}
\title{Plasma instability splits axion-photon conversion lines}

\author{Elizabeth A.~Tolman}

\affiliation{Department of Physics and Astronomy, University of Iowa, Iowa City, IA 52242, USA}
\affiliation{Center for Computational Astrophysics, Flatiron Institute, 162 Fifth Avenue, New York, NY 10010, USA}
\email{etolman@uiowa.edu}

\begin{abstract}
The axion coupling to electromagnetism may leave signatures of uncertain nature in astrophysical spectra; one is thought to be a single line centered at the axion frequency $\omega = \sqrt{\left(m_a c^2/\hbar\right)^2 + c^2 k_a^2}$, with $m_a c^2/\hbar$ the axion Compton frequency and $\hbar k_a$ its momentum. If a parametric plasma instability is excited, this signature is transformed into a \textit{set} of lines, which may be proportionally spaced and linearly polarized. A Crab pulsar signal at a frequency near recent cosmological predictions for the axion Compton frequency has some qualitative similarity to this signature, though this similarity is not evidence that the signal is caused by axions.
\end{abstract}
\maketitle

\section{Introduction}
\label{sec:intro}
The QCD axion offers a potential explanation for  QCD's CP invariance and some or all dark matter \citep{peccei1977constraints,weinberg1978new,kim1979weak,shifman1980can,dine1981simple, dine1983not,abbott1983cosmological,preskill1983cosmology,svrcek2006axions,arvanitaki2010string}, making it---or a similar particle---a candidate for new physics. It may couple to electromagnetism via the Lagrangian term $\mathcal{L}_{a\gamma\gamma} = g_{a\gamma \gamma} a \vec{E} \cdot \vec{B} / \left(4 \pi \right)$, where $g_{a\gamma \gamma}$ is a small coupling constant and $a$, $\vec{E}$, and $\vec{B}$ are the axion, electric, and magnetic fields \citep{wilczek1987two,sikivie2021invisible}. 

This coupling may cause axions in strongly magnetized plasmas to produce an emission signature when they cross ``conversion'' surfaces where the plasma frequency equals the axion Compton frequency $m_a c^2/\hbar$, with $m_a$ the axion mass; the resulting signature is often expected to be a single spectral line at the axion frequency $\omega = \sqrt{\left(m_a c^2/\hbar\right)^2 + c^2 k_a^2}$, with $\hbar k_a$ the axion momentum \citep{lai2006probing,chelouche2008spectral,pshirkov2009conversion,hook2018radio,foster2020green,sikivie2021invisible,battye2022towards,derocco2022first,battye2023searching,an2023searching,todarello2024sun,caputo2024pulsar,noordhuis2024axion,berghaus2025physicsstandardmodeldsa,an2025situ}.

Here,  I show that this  single line splits into a set of lines if a common instability is excited. Specifically, when axions pass a conversion surface in a strongly magnetized plasma, some of them convert into a plasma normal mode similar to the ordinary mode. This is the line thought to be the coupling signature. Nonlinearities create a stimulated-Brillouin-scattering-like parametric instability that could scatter this radiation into other frequencies, causing the axion coupling signature to become a \textit{set} of lines. An instability exactly like the one I describe causes linearly polarized and proportionally spaced lines.  Experiments and simulations show similar instabilities.

The Crab pulsar is surrounded by a strongly magnetized plasma that may have a conversion surface; the pulsar may also host a dense cloud of axions. Its emission has a  feature qualitatively similar to this signature in the range of recent post-inflation QCD axion Compton frequency estimates \citep{buschmann2022dark}, though this similarity is not evidence that the signal is caused by axions.
\section{Model plasma for axion-photon conversion}
\label{sec:stateprob}
This paper characterizes the electromagnetic wave coupling signature resulting from interaction between an axion field and a magnetized plasma with a conversion surface. This interaction might happen in many astrophysical systems, including pulsars, which may be surrounded by strongly magnetized plasmas of spatially varying density and dense clouds of energetic, gravitationally trapped axions originating from the dark matter axion population, magnetospheric electromagnetic processes, and other local physics \citep{goldreich1969pulsar,buschmann2021axion,prabhu2021axion,noordhuis2024axion,caputo2024pulsar}. 
\begin{figure*}[t!]
    \centering
    \includegraphics[width=0.5\textwidth]{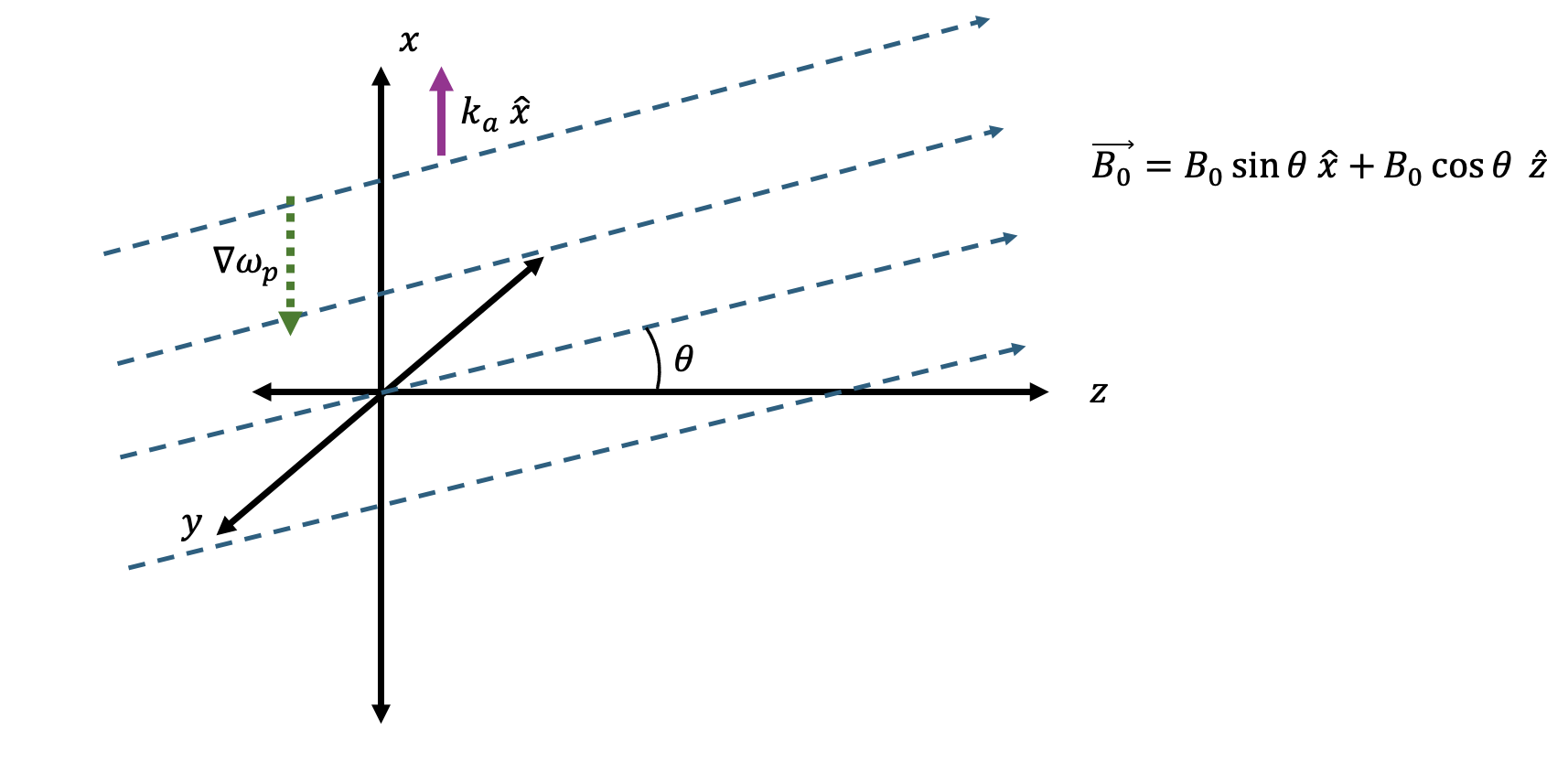}
    \hfill
    \includegraphics[width=0.45\textwidth]{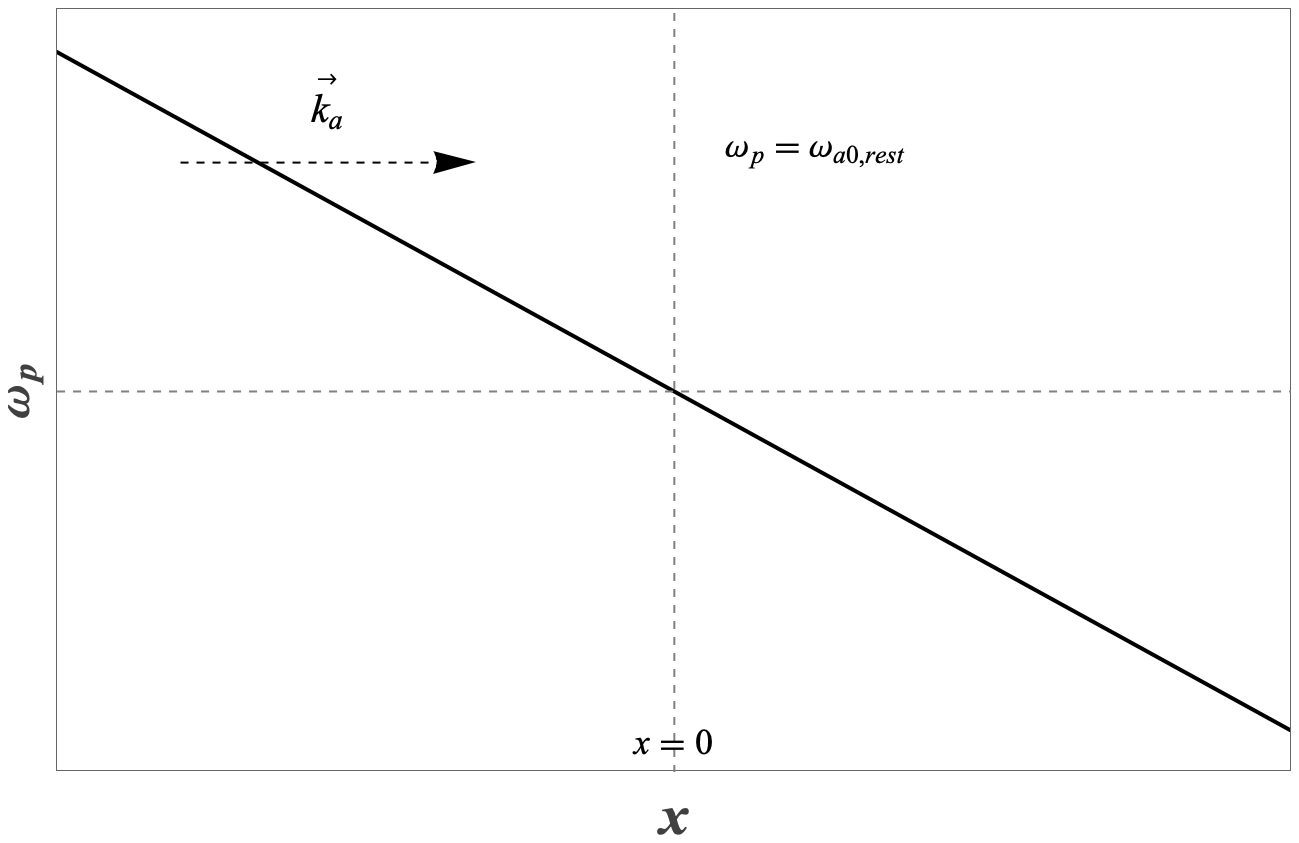}
    \caption{Model configuration for production of axion coupling signature: (left) a strong magnetic field $\vec{B}_0$ lies at an angle $\theta$ to the $\hat{z}$ axis; (right) axions move down a plasma frequency gradient $\nabla \omega_p$ with wave vector $k_a \hat{x}$, producing a coupling signature with parallel wave vector after conversion at $x=0$.}
    \label{fig:tensor}
\end{figure*}

Study of the coupling signature in a specific setting requires computation; to isolate essential physics I study a simplified plasma. This is a quasineutral set of three nonrelativistic, isothermal fluid species $s$: a light, negatively charged electron fluid $s = -$, a light, positively charged positron fluid $s=+$, and a small admixture of a heavy, singly positively charged proton fluid, $s=i$. Its density profile varies with coordinate $x$ about a conversion surface at $x = 0$ where the plasma frequency equals the axion Compton frequency,
\begin{equation}
\label{eq:concond}
\omega_{p} \left(0 \right) = \omega_{a0,rest},
\end{equation}
where the plasma frequency
\begin{equation}
\label{eq:wp}
\omega_{p} \left(x \right) \equiv   \sqrt{ \sum_s \omega_{ps}^2 \left(x \right)},
\end{equation}
with
\begin{equation}
    \omega_{ps}^2 \left(x \right) \equiv \frac{4\pi n_{s,0}\left(x \right) q_s^2}{m_s}, 
\end{equation}
with $n_{s,0}\left(x \right)$,  $q_s$, and $m_{s}$ the background density, charge, and mass of the species $s$ that comprise the plasma, and the axion Compton frequency
\begin{equation}
    \omega_{a0,rest} \equiv \frac{m_a c^2}{\hbar},
\end{equation}
where $m_a$ is the axion mass. Axions approach from $x = -\infty$ and cross the surface with wave vector $k_a \hat{x}$; the plasma wave signature of the axions has wave vectors parallel to the axion wave vector. The plasma has a very strong background magnetic field that lies in the \(x\)-\(z\) plane at a small angle $\theta$ from parallel to the $z$ axis, $\vec{B}_0 = B_0 \sin \theta \hat{x} + B_0 \cos \theta \hat{z}$. Negligible variation in this field provides pressure support. This configuration is pictured in Figure~\ref{fig:tensor}.

The plasma and the axion field are modeled by axion electrodynamics \citep{sikivie1983experimental, wilczek1987two,raffelt1988mixing} and fluid equations. Axion electrodynamics includes a Klein--Gordon equation,
\begin{equation}
\label{eq:afieldeqn}
    \left(\partial_{t}^2 - c^2 \nabla^2 + \omega_{a0,rest}^2 \right) a = \frac{\hbar c^3 g_{a\gamma \gamma}\vec{E} \cdot \vec{B}}{4\pi},
\end{equation}
with $a$ the axion field and $\vec{E}$ and $\vec{B}$ electric and magnetic fields, and modified Maxwell's equations, which combine to form a wave equation,
\begin{multline}
    \label{eq:forwave}
    \partial_{t}^2 \vec{E} + c^2 \nabla \times \left( \nabla \times \vec{E} \right) + 4\pi \partial_t \vec{j}
    + g_{a \gamma \gamma} \partial_{t} \left(\vec{B} \partial_{t} a \right) 
    \\
    + c g_{a \gamma \gamma} \partial_{t} \left( \nabla a \times \vec{E} \right) =0.
\end{multline}
The plasma carries the current    $\vec{j} = \sum_s n_s q_s \vec{v}_s$, with $\vec{v}_s$ the species velocity. Species density and velocity obey fluid equations including a momentum equation,
\begin{equation}
\label{eq:momn}
    m_s n_s \left( \partial_{t} + \vec{v}_s \cdot \nabla \right) \vec{v}_s = n_s q_s \left(\vec{E} + \frac{\vec{v}_s \times \vec{B}}{c}\right)-\nabla p_s,
\end{equation}
where the pressure is    $p_s = m_s c_s^2 n_s$ with $c_s$ the sound speed. I take $c_+= c_-\equiv c_{\pm}$ and assume that $c_\pm \gg c_i$. The equations also include a continuity equation,
\begin{equation}
\label{eq:contn}
    \partial_{t} n_s + \vec{v}_s \cdot \nabla n_s + n_s \nabla \cdot \vec{v}_s=0.
\end{equation}

Plasma quantities consist of prescribed background values and small perturbations of order \(\delta\), where \(\delta\ll1\). These perturbations include the coupling signature produced by the axion field and any other perturbations of the same size from other sources. The axion field (including all axions in the astrophysical system, from all sources)  is $a = \delta a_{1}$; plasma quantities read
\begin{equation}
\label{eq:Bexp}
\begin{aligned}
\phantom{\vec{B}} n_s      &= n_{s,0} + \delta n_{s,1}, \\
\vec{B}                    &= B_{0} \sin \theta \hat{x}+ B_{0} \cos \theta \hat{z} + \delta (B_{1,x}\hat{x}+B_{1,y}\hat{y} + B_{1,z}\hat{z}), \\
\vec{E}                    &= \delta (E_{1,x}\hat{x} + E_{1,y}\hat{y} + E_{1,z}\hat{z}), \\
\phantom{\vec{B}} p_s      &= p_{s,0} + \delta p_{s,1}, \\
\vec{v}_s                  &= \delta \vec{v}_{s,1}, \\
\vec{j}                    &= \vec{j}_{0} + \delta \vec{j}_{1}.
\end{aligned}
\end{equation}
The $\mathcal{O}\left(\delta\right)$ quantities experience fast oscillatory time evolution and slow amplitude evolution due to nonlinear interaction with other $\mathcal{O}\left(\delta\right)$ quantities. Both evolutions are represented by a multiscale expansion \citep{bender2013advanced,kramer2013method} that splits time and space dependence into a fast spatial coordinate $x$ and temporal coordinate $t$ and a slow spatial coordinate $X= \delta x$ and temporal coordinate $T = \delta t$. Denoting a generic $\mathcal{O}\left(\delta\right)$ quantity by $\delta Q_1$, expansions like
\begin{multline}
\label{eq:multsc}
    \delta Q_1 = \delta Q_{1}^{(0)} \left(x,X,t,T \right) 
     + \delta^2 Q_{1}^{(1)} \left(x,X,t,T\right)
    + \mathcal{O}\left(\delta^3\right)
\end{multline}
describe all $\mathcal{O}\left(\delta\right)$ quantities in~\eqref{eq:Bexp}.  

Next, in Sections~\ref{sec:nmode} and~\ref{sec:pdi}, I summarize the wave physics governing the $\mathcal{O}\left(\delta\right)$ quantities. This includes a set of normal modes and conversion between them, given by the $\mathcal{O}\left(\delta\right)$ form of~\eqref{eq:afieldeqn} and~\eqref{eq:forwave}, and a nonlinear instability, given by the $\mathcal{O}\left(\delta^2\right)$ form of~\eqref{eq:forwave}, that results from interaction between the normal modes. Then, in Section~\ref{sec:prop}, I use that physics to describe the axion coupling signature produced in the model plasma system, showing that it can consist of one or multiple lines. Section~\ref{sec:pulsarcloud} qualitatively applies this model to realistic astrophysical systems.

\section{$\mathcal{O}\left(\delta\right)$ normal modes of the axion-plasma system and conversion}
\label{sec:nmode}
The axion coupling signature is determined by~\eqref{eq:afieldeqn} and~\eqref{eq:forwave} evaluated for~\eqref{eq:Bexp}.  In this section, I summarize the physics these equations contain at lowest order, which includes normal-mode oscillation and conversion.

Far from the conversion surface (far from $x = 0$ in Figure~\ref{fig:tensor}), $\mathcal{O}\left(\delta\right)$ quantities oscillate in normal-mode form, with amplitudes that may vary on slow scales. For example, the component of any first-order quantity $\delta Q_1$ that oscillates with frequency $\omega$ is
\begin{multline}
\label{eq:EO1}
\delta Q_1 =
\delta
\left[
\sum_\lambda
Q_{1,\lambda}^{(0)}(X,T)
e^{i(k_\lambda x - \omega t)}
+ \mathrm{c.c.}
\right]
\\
+ \delta^2 Q_1^{(1)}(x,X,t,T)
+ \mathcal{O}(\delta^3).
\end{multline}
Here, \(\lambda\) labels the normal-mode branch and \(k_\lambda(\omega)\)
is the branch's corresponding wave number. Appendix~\ref{sec:dielectric} presents forms of~\eqref{eq:afieldeqn} and~\eqref{eq:forwave} at $\mathcal{O} \left(\delta \right)$  that can be used to find dispersion relations $k_\lambda\left(\omega \right)$. These show the plasma has three nontrivial normal modes: an axion-like normal mode $\lambda = a$, an ordinary-like normal mode $\lambda = O$, and a phonon-like normal mode $\lambda = P$. At lowest nontrivial order in $\theta$, the first two follow from the relationship
\begin{equation}
\label{eq:dispref}
 c^4\left( k_\lambda^2 -k_{a0}^2\right)\left( k_\lambda^2 - k_{O0}^2 \right)  -\frac{\hbar c^3 g_{a\gamma \gamma}^2 B_0^2 \omega^2}{4\pi} = 0,
\end{equation}
where 
\begin{equation}
k_{a0}^2 \left(\omega\right) \equiv \frac{\omega^2 -\omega_{a0,rest}^2}{c^2}, \,
k_{O0}^2 \left(\omega\right) \equiv \frac{\omega^2 -\omega_{p}^2}{c^2}.
\end{equation}
The ``axion-like'' root at $ k_a \approx k_{a0} $ governs a mode that has large excitations of $a_1$ and small excitations of $E_{1,z}$. The ``ordinary-like'' root at \(k_O \approx k_{O0}\) governs a mode that has
large excitations of \(E_{1,z}\) and small excitations of \(a_1\). The
ordinary-like mode perturbs velocity and magnetic field, but not density, with
the following positive-frequency amplitudes:
\begin{align}
v_{s,1,O,z}^{(0)}
&=
\frac{i q_s}{m_s \omega}\,E_{1,O,z}^{(0)}
\label{eq:vO},
\\
B_{1,O,y}^{(0)}
&=
-\frac{c k_O}{\omega}\,E_{1,O,z}^{(0)} .
\label{eq:BO}
\end{align}
The predominantly $\hat{x}$-polarized low-frequency phonon-like mode \citep{nishikawa1968parametric} obeys
\begin{equation}
k_P^2(\omega) \approx
\frac{\omega^2}{\sin^2\theta}
\left[
c_i^2+
\frac{c_\pm^2\omega_{pi}^2}
{\omega_{p-}^2+\omega_{p+}^2+k_P^2c_\pm^2}
\right]^{-1}.
\end{equation}
It features small velocities and densities
\begin{align}
 \vec{v}_{s,1,P}^{(0)}
&=
\frac{i\omega q_s E_{1,P,x}^{(0)}}{m_s\!\left(\omega^2-k_P^2 c_s^2\sin^2\theta\right)}
\left(
\sin^2\theta\,\hat{x}
+\sin\theta\cos\theta\,\hat{z}
\right),
\label{eq:vP}
\\
n_{s,1,P}^{(0)}
&=
\frac{i k_P n_{s,0} q_s E_{1,P,x}^{(0)} \sin^2\theta}{m_s\!\left(\omega^2-k_P^2 c_s^2\sin^2\theta\right)} .
\label{eq:nP}
\end{align}

Near $x=0$, where $\omega_{a0,rest}= \omega_p$, the spatial dependence of $\mathcal{O}\left(\delta\right)$ quantities is not of normal-mode form. Instead, these quantities look like
\begin{multline}
\label{eq:EO1t}
\delta Q_{1} =
\delta
\left[
Q^{(0)}_{1}(x,X,T)
\exp\!\left(-i\omega t\right)
+ \mathrm{c.c.}
\right]
+ \mathcal{O}\left(\delta^2\right),
\end{multline}
where \(Q^{(0)}_{1}(x,X,T)\) is the frequency-domain spatial amplitude. At lowest nontrivial order in $\theta$, 
~\eqref{eq:afieldeqn} and~\eqref{eq:forwave} then reduce to $E_{1,x}^{(0)}=0$ and 
\begingroup
\small
\setlength{\arraycolsep}{2pt}
\begin{multline}
\label{eq:axionE_matrixcon}
\begin{pmatrix}
-\omega^2-c^2 \partial_x^2+\omega_{a0,rest}^2
&
-\frac{\hbar c^3 g_{a\gamma\gamma}B_0}{4\pi}
\\[6pt]
-\omega^2 g_{a\gamma\gamma}B_0
&
-\omega^2 + \omega_p^2\left(x \right)-c^2 \partial_x^2
\end{pmatrix}
\begin{pmatrix}
 a_{1}^{(0)}\\
E_{1,z}^{(0)}
\end{pmatrix}
\\
=0.
\end{multline}
\endgroup
This equation, when solved numerically, shows that, at the conversion surface $x=0$, axion-like and ordinary-like modes mix, allowing conversion from one mode type to another \footnote{This expression can be reduced to Landau--Zener form \citep{Landau1932,Zener1932,rubbmark1981dynamical}.}. This conversion is discussed more in Appendix~\ref{sec:con}.

\section{$\mathcal{O}\left(\delta^2 \right)$ nonlinear parametric instability}
\label{sec:pdi}
Here, I summarize the \(\mathcal{O}(\delta^2)\) wave physics, which describes nonlinear interaction that causes normal-mode
excitations of the plasma to experience slow length- and time-scale amplitude
evolution \citep{su1994nonlinear,bingham1996nonlinear,gomberoff1997parametric,
mendoncca2007axion,shi2019three,edwards2016strongly,beyer2023parametric,
nishiura2025unified, hook2025cosmologicalconstraintsdarkphoton}. 

I discuss how
this physics affects an individual ordinary-like wave, labeled \(O_1\), with
frequency \(\omega_{O_1}\), wave number \(k_{O_1}\), and amplitude
\(E_{1,O_1,z}^{(0)}\), far from the conversion surface, where its axion field
component is negligible. The plasma that hosts this wave is rippled with
seed-level excitations of the complete spectrum of possible ordinary-like and
phonon-like waves. Two sets of these modes are special. A Stokes set obeys
\begin{equation}
\label{eq:3ww}
    \omega_{O_1}  = \omega_{O_2} +\omega_{P_1}, \,
    k_{O_1}  = k_{O_2} + k_{P_1},
\end{equation}
with 
frequency \(\omega_{O_2}\) and wave number \(k_{O_2}\) characterizing an ordinary-like seed wave \(O_2\), and frequency \(\omega_{P_1}\) and
wave number \(k_{P_1}\) characterizing a
phonon-like seed wave \(P_1\). A corresponding anti-Stokes set obeys
\begin{equation}
\label{eq:3ww2}
   \omega_{O_1} + \omega_{P_2} = \omega_{O_3}, \,
    k_{O_1} + k_{P_2} = k_{O_3},
\end{equation}
with $O_3$ a third ordinary-like wave and $P_2$ another phonon-like wave.

The Stokes set participates in a feedback loop that results in an instability causing the seed waves
\(O_2\) and \(P_1\) to increase in amplitude. The initial ordinary-like wave's
velocity perturbation \(v_{s,1,O_1,z}\)~\eqref{eq:vO} and the seed
phonon-like wave's density perturbation \( n_{s,1,P_1}\)~\eqref{eq:nP}
cause a beat current at \(k_{O_2}\) and \(\omega_{O_2}\) that drives the
ordinary-like wave \(O_2\). This wave's and the initial ordinary-like wave's
velocity and magnetic-field perturbations,~\eqref{eq:vO} and~\eqref{eq:BO}, create beat forces at \(k_{P_1}\) and \(\omega_{P_1}\),
amplifying the interaction. Phonon-like perturbations that satisfy the anti-Stokes resonance
condition~\eqref{eq:3ww2} can similarly scatter the ordinary-like
wave \(O_1\) into a higher-frequency ordinary-like wave \(O_3\). The ordinary-like waves driven in these
interactions are determined by~\eqref{eq:3ww},~\eqref{eq:3ww2}, and the
dispersion relations of the ordinary-like and phonon-like branches.

In Appendix~\ref{sec:pdideriv}, I derive that the instability's growth is given by 
\begin{equation}
\label{eq:inst}
     E_{1,O_2,z}^{(0)} \left(X,T\right) = E_{1,O_2,z}^{(0)}\left(X, 0\right) e^{\Gamma T },
\end{equation}
with 
\begin{equation}
\label{eq:pdigrow}
   \Gamma^2 =  
  \kappa_{O_1,P_1} \kappa_{O_1,O_2} \left| E_{1,O_1,z}^{(0)}\right|^2 .
\end{equation}
The coupling constants,  $\kappa_{O_1,P_1}$ and $\kappa_{O_1,O_2}$, defined in Appendix~\ref{sec:pdideriv}, have magnitudes
estimated for typical pulsar parameters in Table~\ref{tab:pdi_scalings} of Appendix~\ref{sec:typ}. 

The instability is damped by spatial inhomogeneity limiting the region where
the three waves obey~\eqref{eq:3ww} \citep{rosenbluth1972parametric,kruer2019physics}. To model this inhomogeneity, I
promote wave numbers of the modes involved in the parametric instability to
spatially dependent quantities (assuming geometric optics) and define
\begin{equation}
\varkappa(x) \equiv
k_{O_1}(x)
-
k_{P_1}(x)
-
k_{O_2}(x)
\end{equation}
for condition~\eqref{eq:3ww}. Then, the criterion for the first ordinary-like wave to
drive the phonon-like wave and the second ordinary-like wave in spite of inhomogeneity is
\begin{equation}
\label{eq:dampeq}
    \frac{\delta^2\kappa_{O_1,P_1} \kappa_{O_1,O_2}
    \left|E_{1,O_1,z}^{(0)}\right|^2}
    {\left|\partial_{x} \varkappa\right| \left| v_{g,O_2} v_{g,P_1} \right|}
    \gtrsim 1 ,
\end{equation}
with $v_{g,O_2}$ and $ v_{g,P_1}$ the  group velocities of the second ordinary-like and the phonon-like modes. 
For typical pulsar magnetosphere conditions, the size of the left-hand side of this equation is estimated
in Appendix~\ref{sec:typ}, with the results shown in Figure~\ref{fig:pdiparam}; the estimate shows that the instability is easy to excite.

The ordinary-like waves driven by these interactions can participate in additional three-wave interactions, producing a cascade of ordinary-like modes. The frequencies of these modes are roughly proportionally spaced. To show this, I consider two counterpropagating ordinary-like modes with $\omega_{O_1}, \omega_{O_2} \gg \omega_{P_1}$, $k_{O_1} \approx -k_{O_2}$, so $|k_{P_1}| \approx 2 |k_{O_2}|$. Then, with the spacing  given by $\Delta \omega \equiv \omega_{O_1} - \omega_{O_2}$, Eq.~\eqref{eq:3ww} gives 
\begin{multline}
\label{eq:propspace}
    \Delta \omega
    =
    2\sin\theta\, |k_{O_2}|
    \left[
    c_i^2
    +
    \frac{c_\pm^2\,\omega_{pi}^2}
    {\omega_{p-}^2+\omega_{p+}^2+4k_{O_2}^2c_\pm^2}
    \right]^{1/2}
    \\
    \approx
    2\sin\theta\, \omega_{O_2}
    \left[
    \frac{c_i^2}{c^2}
    +
    \frac{c_\pm^2}{c^2}
    \frac{\omega_{pi}^2}
    {\omega_{p-}^2+\omega_{p+}^2+4\omega_{O_2}^2c_\pm^2/c^2}
    \right]^{1/2}.
\end{multline}
The corresponding normalized estimate is given in the final row of
Table~\ref{tab:pdi_scalings}.

\section{Spectral coupling signature produced by propagation of axions across conversion surface} 
\label{sec:prop}
Now, I use the physics developed in previous sections to determine the coupling signature in the model configuration (pictured in Figure~\ref{fig:tensor}), which retains the basic qualities of astrophysical systems where axion conversion might occur. Axions approaching the conversion surface
from \(x<0\) are described by the axion-like root of the
dispersion relation~\eqref{eq:dispref}. When this axion wave traverses the
conversion surface, mode conversion physics in~\eqref{eq:axionE_matrixcon} converts part of it into an ordinary-like wave; I discuss this conversion in Appendix~\ref{sec:con}. This
ordinary-like wave is the single-line axion-photon conversion signature usually
considered in plasma-based searches.

However, as Section~\ref{sec:pdi} shows, such an ordinary-like wave pumps a
parametric instability if~\eqref{eq:dampeq} holds. The result is not a single isolated wave, but a
cascade of ordinary-like sidebands formed by waves satisfying~\eqref{eq:3ww} and~\eqref{eq:3ww2}. These sidebands remain linearly polarized
and are proportionally spaced in frequency, as modeled by~\eqref{eq:propspace}. 

\begin{figure}[t]
    \centering
    \includegraphics[width=\columnwidth]{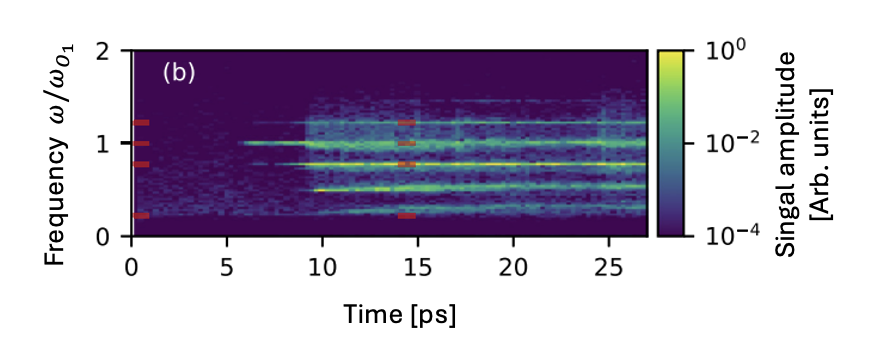}\\[1ex]
    \includegraphics[width=\columnwidth]{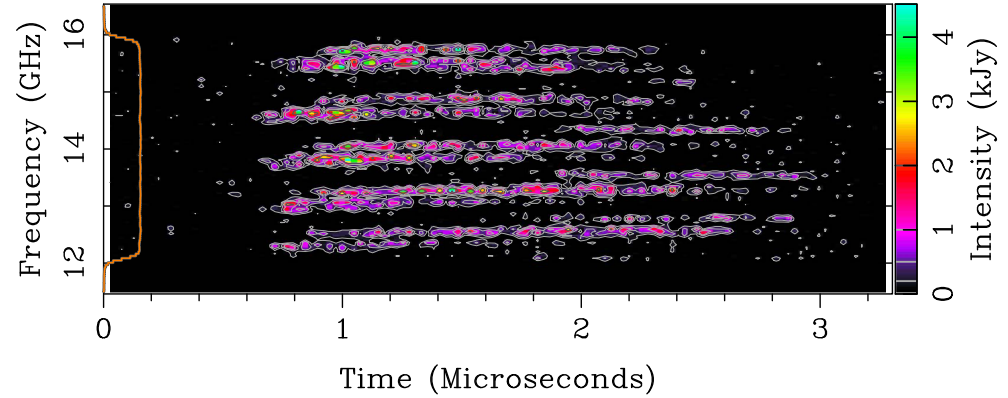}
    \caption{
    (Top) Simulation of an initially monochromatic plasma wave of frequency
    \(\omega_{O_1}\), analogous to the ordinary-like wave from axion conversion,
    moving through plasma and producing a set of emission lines through
    parametric instability \citep{blackman2024impact}. (Bottom) Observed
    emission signature in the Crab pulsar \citep{hankins2016crab}. (© AAS;
    reproduced with permission.) The emission bands are nearly completely
    linearly polarized and proportionally spaced. Doubled emission bands may result
    from the same process occurring in regions of plasma with different
    velocities and Doppler shifts.
    }
    \label{fig:crab}
\end{figure}

Similar processes are observed in ionospheric plasmas, laboratory plasma experiments, and simulations, where the general spectral appearance is not very sensitive to the conditions of the underlying plasma 
\citep{stenzel1975parametric,fejer1979ionospheric,li2001observation,li2012characterization,
hansen2019parametric,hansen2020parametric,wolff2021brillouin,blackman2024impact}.
One example is the top panel of Figure~\ref{fig:crab}, which shows an initially
monochromatic wave transformed into a set of emission lines as it moves through
a plasma.

The axion-photon coupling
signature in plasma could be a similar set of lines rather than a single line. The specific wave coupling I describe in this paper yields a completely linearly polarized,
proportionally spaced set, but an instability resulting from a different wave coupling may have different properties.  Recent post-inflation QCD axion mass estimates suggest an axion Compton frequency $\omega_{a0,rest}/ \left(2 \pi \right)$ near \(9\)--\(44\,{\rm GHz}\) \citep{buschmann2022dark}, meaning that this pattern could be centered at a frequency in or above this range, depending on the momentum of the axions that seed the pattern.

\section{Qualitative application to pulsar signal}
\label{sec:pulsarcloud}

It is interesting to ask if any known astrophysical signals
resemble the spectral pattern described above. This investigation is qualitative only. Astrophysical systems have geometric, relativistic, and streaming effects not considered in this paper; in addition, I cannot distinguish axion-induced emission from standard plasma astrophysics.

A natural place to look for qualitatively similar spectral patterns is the Crab pulsar, a bright, well-observed,
strongly magnetized neutron star that has been discussed as a possible efficient
source of axions produced by electromagnetic processes in its magnetosphere
\citep{khelashvili2025axion}.  Some of these axions remain trapped around the star as a cloud. These axions' initial momenta, set by the pair discharge in which they are produced, may be dominated by a characteristic momentum at the surface of the star satisfying roughly
\citep{caputo2024pulsar}
\begin{equation}
\label{eq:source}
    \hbar c k_{a}\left(r = r_*\right)   \approx \frac{m_a c^2}{3},
\end{equation}
with $r$ the radius from the center of the star and $r_*$ the radius at the stellar surface;
this characteristic momentum then varies according to
\begin{equation}
\label{eq:momrel}
   \frac{\hbar^2 k_{a}^2\left(r\right)}{m_a^2}
    =
   \frac{\hbar^2 k_{a}^2\left(r_* \right) }{m_a^2}
   - 2GM\left(\frac{1}{r_*}-\frac{1}{r}\right),
\end{equation}
with $M$ the mass of the pulsar. These axions traverse the plasma that surrounds the neutron star, encountering a spatially and temporally varying plasma frequency $\omega_p\left(r, \vartheta,\phi, t\right)$ \citep{tjemsland2024adiabatic,satherley2025axion}, with $\vartheta$ and $\phi$ polar and azimuthal angles. If this plasma frequency profile includes a surface where~\eqref{eq:concond} holds, cloud axions could convert into ordinary-like waves with a width in frequency space determined by the spread of source axions around~\eqref{eq:source} and the location of the conversion surface \citep{caputo2024pulsar,berghaus2025physicsstandardmodeldsa}.
These could pump the parametric
instability discussed in Section~\ref{sec:pdi}.

In fact, the Crab displays a striking radio feature with several qualitative
similarities to the pattern discussed in Section~\ref{sec:prop}. In particular, its high-frequency interpulse contains a
remarkably stable set of emission bands, sometimes called a ``zebra pattern''
\citep{hankins2007radio,hankins2016crab,medvedev2024origin} that begin
at an observed frequency of about \(5\,{\rm GHz}\) and extend to at least
\(30\,{\rm GHz}\) \footnote{Observed frequencies are Doppler shifted relative to source frequencies; see
\cite{zheleznyakov2012analogy}.}. The bands are nearly completely linearly
polarized and approximately proportionally spaced with $\Delta \omega = 0.06 \omega$
\citep{hankins2016crab,medvedev2024origin}. Proportional spacing is also seen in the parametric
cascade described by Eq.~\eqref{eq:propspace}. The appearance of the signal suggests a parametric instability pumped by a strong, narrow-band,
linearly polarized wave, though it is not possible at this point to determine the source of the pump.

The Crab zebra feature has other unusual features, including a constant position angle and a dispersion measure that is variable and often larger than that of typical radio emission \citep{hankins2007radio,hankins2016crab,medvedev2024origin}. The first of these characteristics is not explained by axion-induced conversion unless it results from some unknown symmetry of the conversion surface or the axion cloud. The second could have some relationship to enhanced density structure induced by the phonon-like mode.

I note that similar zebra-like emission patterns also occur in solar radio bursts, despite the very
different plasma environment
\citep{zheleznyakov2012analogy,karlicky2013radio,chernov2015latest}, showing that sets of narrow radio bands may occur more broadly in nature.  

\section{Conclusion}
\label{sec:concl}
I analyzed the linear and nonlinear interaction of an axion field with a magnetized plasma to determine the emission signature produced by axions moving across a conversion surface. The axions, described by an axion-like normal mode, partially convert into an ordinary-like linear normal mode when they cross the conversion surface. This ordinary-like wave may drive a nonlinear parametric instability, which transforms the single-line coupling signature into a multiple-line coupling signature. The class of instability described here naturally produces linearly polarized, proportionally spaced sidebands. Laboratory experiments and simulations show similar spectra. Crab pulsar data exhibit qualitatively similar spectral features, though this similarity is not evidence that the features are caused by axions.

Future work should use computational methods to extend the present treatment to conditions that more precisely resemble astrophysical systems.  In particular, these efforts should consider how the instability occurs in relativistic and streaming plasmas \citep{tolman2022electric}. Additional work should also consider effects that can reduce damping \citep{white1974amplification,hansen2020parametric}. The analysis should be extended to apply to the spectral signature of other beyond-the-Standard-Model candidates, like the dark photon \citep{caputo2021dark}.  Then, additional analysis can be conducted on astrophysical data.

\begin{acknowledgments}
I am grateful for helpful conversations with Olivier Simon, Ani Prabhu, Misha Medvedev, Hayk Hakobyan, Chris Dessert, Sasha Philippov, Ilya Dodin, and Sam Witte. I am also grateful for helpful comments from two anonymous reviewers. GPT-5.6 Sol checked algebraic calculations conducted by the author and proofread the manuscript, offering comments that were reviewed by the author as though they were from a colleague. Research at the Flatiron Institute is supported by the Simons Foundation. This research was supported in part by grant NSF PHY-2309135 to the Kavli Institute for Theoretical Physics (KITP).  Figures are used with permission.
\end{acknowledgments} 

\appendix

\section{Derivation of wave equation for setup in text}
\label{sec:dielectric}
This text concerns the propagation of waves along $\hat{x}$ through an oblique magnetic field at a
small angle to $\hat{z}$, $\vec{B}_0 = B_0 \sin \theta \,\hat{x} + B_0 \cos \theta\, \hat{z}$,
as illustrated in Figure~\ref{fig:tensor}. Here, I derive the wave equation for this configuration. The wave equations~\eqref{eq:afieldeqn} and~\eqref{eq:forwave} at $\mathcal{O}\left(\delta\right)$ read
\begin{equation}
\label{eq:wave_Odelta}
\left(\partial_t^2 - c^2 \partial_x^2 + \omega_{a0,rest}^2 \right) a_1^{(0)}
= \frac{\hbar c^3 g_{a\gamma\gamma}\,\vec E_1^{(0)} \cdot \vec B_0 }{4\pi}
\end{equation}
and
\begin{multline}
\label{eq:wave_Odelta2}
\partial_t^2 \vec E_1^{(0)}
+c^2\nabla_x\times(\nabla_x\times \vec E_1^{(0)})
+4\pi \partial_t \vec j_1^{(0)}
\\
+g_{a\gamma\gamma}\partial_t\!\left(\vec B_0\,\partial_t a_1^{(0)}\right)
=0 .
\end{multline}
For fields of form~\eqref{eq:EO1}, Eqs.~\eqref{eq:wave_Odelta} and~\eqref{eq:wave_Odelta2} become
\begin{equation}
\left(-\omega^2 + c^2 k_\lambda^2 + \omega_{a0,rest}^2 \right)
a_{1,\lambda}^{(0)}
=
\frac{\hbar c^3 g_{a\gamma\gamma}\,
\vec E_{1,\lambda}^{(0)} \cdot \vec{B}_0}{4\pi}
\end{equation}
and
\begin{multline}
k_\lambda \hat{x} \times
\left(k_\lambda \hat{x} \times \vec E_{1,\lambda}^{(0)}\right)
+\frac{\omega^2}{c^2}\,\overleftrightarrow{\epsilon}_{\lambda}
\cdot \vec E_{1,\lambda}^{(0)}
\\
+ \frac{g_{a\gamma\gamma} \omega^2 \vec{B}_0 a_{1,\lambda}^{(0)}}{c^2}
=0,
\end{multline}
where the dielectric tensor for the $\lambda$th mode component is defined by
\begin{equation}
\overleftrightarrow{\epsilon}_{\lambda}
=
\overleftrightarrow{I}
+
\sum_s \overleftrightarrow{\chi}_{s,\lambda}.
\end{equation}
Here, the susceptibility of species $s$ evaluated at $(k_\lambda,\omega)$ may be written as
\begin{widetext}
\begin{equation}
\begin{aligned}
\overleftrightarrow{\chi}_{s,\lambda}
&=
-\frac{\omega_{ps}^2}{\Delta_{s,\lambda}}
\\[4pt]
&\quad\times
\begin{pmatrix}
\omega^2-\Omega_s^2\sin^2\theta &
i\omega\Omega_s\cos\theta &
-\Omega_s^2\sin\theta\cos\theta
\\[1mm]
-i\omega\Omega_s\cos\theta &
\Lambda_{s,\lambda} &
i\Lambda_{s,\lambda}\Omega_s\sin\theta/\omega
\\[1mm]
-\Omega_s^2\sin\theta\cos\theta &
-i\Lambda_{s,\lambda}\Omega_s\sin\theta/\omega &
\Lambda_{s,\lambda}-\Omega_s^2\cos^2\theta
\end{pmatrix},
\end{aligned}
\end{equation}
\end{widetext}
where \(\Omega_s\equiv q_s B_0/(m_s c)\) is the signed cyclotron frequency,
and
\begin{equation}
\begin{aligned}
\Lambda_{s,\lambda} &= \omega^2-k_\lambda^2c_s^2, \\
\Delta_{s,\lambda} &=
\Lambda_{s,\lambda}(\omega^2-\Omega_s^2\sin^2\theta)
-\omega^2\Omega_s^2\cos^2\theta .
\end{aligned}
\end{equation}
In the strongly magnetized limit, where $\omega, \, \omega_{ps}, \, k_\lambda^2 c_s^2/ \omega \ll \left|\Omega_s\right|$, this susceptibility reduces to
\begin{equation}
\overleftrightarrow{\chi}_{s,\lambda}
\approx
-\frac{\omega_{ps}^2}{\omega^2-k_\lambda^2 c_s^2\sin^2\theta}
\begin{pmatrix}
\sin^2\theta & 0 & \sin\theta\cos\theta
\\[1mm]
0 &
0 &
0
\\[3mm]
\sin\theta\cos\theta &
0 &
\cos^2\theta
\end{pmatrix}.
\end{equation}
In this limit the \(y\)-polarized electromagnetic mode is not interesting, so when considering the full set of wave equations I restrict attention to the axion-\(x\)-\(z\) block. Then, for a given mode component $\lambda$, Eqs.~\eqref{eq:wave_Odelta} and~\eqref{eq:wave_Odelta2} can be stated as
\begin{equation}
\label{eq:axionE_matrix_compact}
\omega^2
\begin{pmatrix}
\Delta_{a,\lambda} & \eta_{x,\lambda} & \eta_{z,\lambda} \\
\beta_{x,\lambda}  & \epsilon_{xx,\lambda} & \epsilon_{xz,\lambda} \\
\beta_{z,\lambda}  & \epsilon_{xz,\lambda} &
\epsilon_{zz,\lambda}-\dfrac{c^2k_\lambda^2}{\omega^2}
\end{pmatrix}
\begin{pmatrix}
a_{1,\lambda}^{(0)} \\
E_{1,\lambda,x}^{(0)} \\
E_{1,\lambda,z}^{(0)}
\end{pmatrix}
=0,
\end{equation}
where the values in the matrix are defined in Table~\ref{tab:aE}.
\begin{table}[h!]
\centering
\begin{tabular}{ll}
\toprule
\textbf{Quantity} & \textbf{Definition} \\
\midrule
$\Delta_{a,\lambda}$ 
& $-1+\dfrac{c^2k_\lambda^2}{\omega^2}+\dfrac{\omega_{a0,rest}^2}{\omega^2}$ \\[8pt]

$\eta_{x,\lambda}$ 
& $-\dfrac{\hbar c^3 g_{a\gamma\gamma} B_0 \sin\theta}{4 \pi \omega^2}\,$ \\[8pt]

$\eta_{z,\lambda}$ 
& $-\dfrac{\hbar c^3 g_{a\gamma\gamma} B_0 \cos\theta}{4 \pi \omega^2}\,$ \\[8pt]

$\beta_{x,\lambda}$ 
& $g_{a\gamma\gamma} B_0 \sin\theta$ \\[8pt]

$\beta_{z,\lambda}$ 
& $g_{a\gamma\gamma} B_0 \cos\theta$ \\[8pt]

$\epsilon_{xx,\lambda}$ 
& $1 - \sin^2\theta \displaystyle\sum_s 
\frac{\omega_{ps}^2}{\omega^2 - k_\lambda^2 c_s^2 \sin^2\theta}$ \\[10pt]

$\epsilon_{xz,\lambda}$ 
& $-\sin\theta \cos\theta \displaystyle\sum_s 
\frac{\omega_{ps}^2}{\omega^2 - k_\lambda^2 c_s^2\sin^2\theta}$ \\[10pt]

$\epsilon_{zz,\lambda}$ 
& $1 - \cos^2\theta \displaystyle\sum_s 
\frac{\omega_{ps}^2}{\omega^2 - k_\lambda^2 c_s^2\sin^2\theta}$ \\
\bottomrule
\end{tabular}
\caption{Definitions of coefficients and dielectric tensor components in Eq.~\eqref{eq:axionE_matrix_compact}.}
\label{tab:aE}
\end{table}
The pressure-driven, phonon-like branch is determined at leading order by
\begin{equation}
\label{eq:presdis}
\begin{aligned}
\epsilon_{xx,P}
&=
1-\sin^2\theta\sum_s
\frac{\omega_{ps}^2}{\omega^2-k_P^2 c_s^2\sin^2\theta}
\\
&\approx 0.
\end{aligned}
\end{equation}

\section{Parametric instability}
\label{sec:pdideriv}

The ordinary and pressure-driven modes discussed in Section~\ref{sec:nmode} participate in a parametric instability discussed qualitatively in Section~\ref{sec:pdi}. Here, I quantitatively derive the growth rate of this instability in the region where the axion field can be neglected. At order $\mathcal{O}(\delta^2)$, the wave equation~\eqref{eq:forwave} becomes
\begin{multline}
\label{eq:wave_Odelta2a}
\partial_t^2 \vec E_1^{(1)}
+c^2\nabla_x\times(\nabla_x\times \vec E_1^{(1)})
+2\partial_t\partial_T \vec E_1^{(0)}
\\
+c^2\!\left[
\nabla_x\times(\nabla_X\times\vec E_1^{(0)})
+\nabla_X\times(\nabla_x\times\vec E_1^{(0)})
\right]
\\
+4\pi\!\left(\partial_t \vec j_1^{(1)}+\partial_T \vec j_1^{(0)}\right)
=0,
\end{multline}
with
\begin{equation}
\begin{aligned}
\vec j_1^{(0)}
&=\sum_s q_s n_{s,0}\vec v_{s,1}^{(0)}, \\
\vec j_1^{(1)}
&=\sum_s q_s\!\left(n_{s,0}\vec v_{s,1}^{(1)}
+n_{s,1}^{(0)}\vec v_{s,1}^{(0)}\right).
\end{aligned}
\end{equation}
Now, I consider the three waves discussed in Section~\ref{sec:pdi}: an ordinary-like wave excited by a physical process, labeled $O_1$, with wave number $k_{O_1}$, frequency $\omega_{O_1}$, and amplitude $E_{1,O_1,z}^{(0)}(X,T)$; a seed-level second ordinary-like wave, labeled $O_2$, with wave number $k_{O_2}$, frequency $\omega_{O_2}$, and amplitude $E_{1,O_2,z}^{(0)}(X,T)$; and a phonon-like wave, labeled $P_1$, with wave number $k_{P_1}$, frequency $\omega_{P_1}$, and amplitude $E_{1,P_1,x}^{(0)}(X,T)$. These waves obey condition~\eqref{eq:3ww}.

To obtain an envelope equation governing the slow time evolution of the second ordinary-like wave participating in Eq.~\eqref{eq:3ww}, I extract from the second-order wave equation the component proportional to $\exp[i(k_{O_2}x-\omega_{O_2}t)]$.
The resonant part of the \(z\)-component then satisfies
\begin{equation}
\begin{aligned}
&\left[
\left(\partial_t^2-c^2\partial_x^2+\omega_p^2\right)
E_{1,z}^{(1)}
\right]_{O_2,\mathrm{res}}
\\
&\quad=
\left[
2\left(-\partial_{tT}^2+c^2\partial_{xX}^2\right)
E_{1,O_2,z}^{(0)}
\right]
\\
&\qquad-
4\pi \sum_s q_s
\left(n_{s,1,P_1}^{(0)}\right)^*
\partial_t v_{s,1,O_1,z}^{(0)}
\\
&\qquad-
4\pi \sum_s q_s
v_{s,1,O_1,z}^{(0)}
\partial_t\left(n_{s,1,P_1}^{(0)}\right)^* .
\end{aligned}
\end{equation}
Here, \([\cdots]_{O_2,\mathrm{res}}\) denotes the component with fast phase $\exp[i(k_{O_2}x-\omega_{O_2}t)]$.

For the correction \(E_{1,z}^{(1)}\) to remain bounded, the coefficient of \(\exp[i(k_{O_2}x-\omega_{O_2}t)]\) on the right-hand side of the equation must vanish.
This solvability condition yields, for condition~\eqref{eq:3ww},
\begin{equation}
\label{eq:varyO_2}
\begin{aligned}
\left( \partial_{T}+v_{g,O_2}\partial_{X} \right)
E_{1,O_2,z}^{(0)}
&={}
\\[-2pt]
&\quad-\kappa_{O_1,P_1}E_{1,O_1,z}^{(0)}
\left(E_{1,P_1,x}^{(0)}\right)^{*},
\end{aligned}
\end{equation}
where $\kappa_{O_1,P_1}$ is
\begin{equation}
\label{eq:kappaO1P_exact}
\begin{aligned}
\kappa_{O_1,P_1}
&=
\frac{k_{P_1}}{2\omega_{O_1}}
\sum_s \frac{q_s}{m_s}
\frac{\omega_{ps}^2\sin^2\theta}
{\omega_{P_1}^2-k_{P_1}^2c_s^2\sin^2\theta}
\\
&\approx
\frac{e}{2\omega_{O_1}k_{P_1}c_\pm^2}
\left[
\begin{aligned}
&\frac{\omega_{p-}^2+\omega_{p+}^2+k_{P_1}^2c_\pm^2}{m_i}
\\[-2pt]
&\quad+\frac{\omega_{p-}^2-\omega_{p+}^2}{m_e}
\end{aligned}
\right],
\end{aligned}
\end{equation}
and the second ordinary-like group velocity is $v_{g,O_2}= c^2 k_{O_2}/\omega_{O_2}$; the complex conjugate equation holds for $\left(E_{1,O_2,z}^{(0)}\right)^{*}$. To derive this expression, I used~\eqref{eq:vO} and~\eqref{eq:nP}. For condition~\eqref{eq:3ww2}, the corresponding envelope equation is
\begin{equation}
\left(\partial_T+v_{g,O_3}\partial_X\right)
E_{1,O_3,z}^{(0)}
=
+\kappa_{O_1,P_2}
E_{1,O_1,z}^{(0)}
E_{1,P_2,x}^{(0)},
\end{equation}
where \(\kappa_{O_1,P_2}\) is given by Eq.~\eqref{eq:kappaO1P_exact} with \(P_1\) replaced by \(P_2\).
These expressions describe variation on long length and time scales in the second and third ordinary-like waves caused by small currents at their frequencies from the beat of the corresponding phonon-like and first ordinary-like wave perturbations.

The phonon-like wave envelope equation may be obtained directly from the
Manley--Rowe relation \citep{manley1956some}, which expresses conservation of wave action in resonant
three-wave interactions, and Eq.~\eqref{eq:varyO_2}. For the matching condition~\eqref{eq:3ww}, the Manley--Rowe relation reads
\begin{equation}
\begin{aligned}
-&\left(\partial_T+v_{g,O_1}\partial_X\right)
\left(\frac{\mathcal{E}_{O_1}}{\omega_{O_1}}\right)
\\
&=\left(\partial_T+v_{g,P_1}\partial_X\right)
\left(\frac{\mathcal{E}_{P_1}}{\omega_{P_1}}\right)
\\
&=\left(\partial_T+v_{g,O_2}\partial_X\right)
\left(\frac{\mathcal{E}_{O_2}}{\omega_{O_2}}\right),
\end{aligned}
\end{equation}
where \(\mathcal{E}_j\) is the energy density of wave $j$,
\begin{equation}
\begin{aligned}
\mathcal{E}_j
&= \frac{\omega_j}{4\pi}
\left.\left(\frac{\partial D_{\lambda_j}}{\partial\omega}\right)
\right|_{\omega=\omega_j,\,k=k_j}
\\
&\qquad\times\left|E_{1,j}^{(0)}\right|^2,
\end{aligned}
\end{equation}
where $D_{\lambda_j}$ is the dimensionless dispersion relation of the normal mode that the wave excites (i.e., $\lambda_{P_1} = P$, etc.) and $E_{1,j}^{(0)}$ is the electric field amplitude of wave $j$. Using Eq.~\eqref{eq:varyO_2}, multiplying by the complex conjugate, adding the conjugate equation, and
substituting into the Manley--Rowe relation gives
\begin{multline}
\left.\left(\frac{\partial D_{P}}{\partial\omega}\right)
\right|_{\omega=\omega_{P_1},\,k=k_{P_1}}
\left(\partial_T+v_{g,P_1}\partial_X\right)
\left|E_{1,P_1,x}^{(0)}\right|^2
\\
=-\frac{2\kappa_{O_1,P_1}}{\omega_{O_2}}
\left[
\left(E_{1,O_1,z}^{(0)}\right)^{*}
E_{1,P_1,x}^{(0)}E_{1,O_2,z}^{(0)}
\right.
\\
\left.
+E_{1,O_1,z}^{(0)}
\left(E_{1,P_1,x}^{(0)}\right)^{*}
\left(E_{1,O_2,z}^{(0)}\right)^{*}
\right].
\end{multline}
This gives the envelope equation
\begin{multline}
\label{eq:varyP}
\left(\partial_T+v_{g,P_1}\partial_X\right)
\left(E_{1,P_1,x}^{(0)}\right)^{*}
\\
=-\kappa_{O_1,O_2}
\left(E_{1,O_1,z}^{(0)}\right)^{*}
E_{1,O_2,z}^{(0)},
\end{multline}
where
\begin{equation}
\label{eq:kappaO1O2_def}
\begin{aligned}
\kappa_{O_1,O_2}
&=
\frac{2\kappa_{O_1,P_1}}
{\omega_{O_2}}
\\
&\quad\times
\left(\frac{\partial D_P}{\partial\omega}\right)^{-1}_{\omega=\omega_{P_1},\,k=k_{P_1}}.
\end{aligned}
\end{equation}
Eq.~\eqref{eq:presdis} gives
\[
\begin{aligned}
\left.\left(\frac{\partial D_P}{\partial\omega}\right)
\right|_{\omega=\omega_{P_1},\,k=k_{P_1}}
&=2\omega_{P_1}\sum_s
\\[-2pt]
&\quad\times
\frac{\omega_{ps}^2\sin^2\theta}
{\left(\omega_{P_1}^2-k_{P_1}^2c_s^2\sin^2\theta\right)^2},
\end{aligned}
\]
such that
\begin{equation}
\label{eq:kappaO1O_2_exact}
\begin{aligned}
\kappa_{O_1,O_2}
&=
\frac{\kappa_{O_1,P_1}}
{\omega_{O_2}\omega_{P_1}
\displaystyle\sum_s
\frac{\omega_{ps}^2\sin^2\theta}
{\left(\omega_{P_1}^2-k_{P_1}^2c_s^2\sin^2\theta\right)^2}}
\\[2pt]
&\approx
\frac{e k_{P_1}^3c_\pm^2\sin^2\theta}
{2\omega_{O_1}\omega_{O_2}\omega_{P_1}}
\\[-2pt]
&\quad\times
\frac{
\displaystyle
\left[
\frac{\omega_{p-}^2+\omega_{p+}^2+k_{P_1}^2c_\pm^2}{m_i}
+\frac{\omega_{p-}^2-\omega_{p+}^2}{m_e}
\right]
}{
\displaystyle
\left[
\omega_{p-}^2+\omega_{p+}^2
+\frac{\left(\omega_{p-}^2+\omega_{p+}^2+k_{P_1}^2c_\pm^2\right)^2}{\omega_{pi}^2}
\right]
}.
\end{aligned}
\end{equation}

Expressions~\eqref{eq:varyO_2} and~\eqref{eq:varyP} can be used to derive a differential equation governing the growth of the three-wave interaction discussed in Section~\ref{sec:pdi}. Taking all fields to be roughly spatially uniform and neglecting depletion of the pump ordinary-like wave \citep{hansen2019parametric} so that
\begin{equation}
\left|
E_{1,P_1,x}^{(0)}
\partial_T E_{1,O_1,z}^{(0)}
\right|
\ll
\left|
E_{1,O_1,z}^{(0)}
\partial_T E_{1,P_1,x}^{(0)}
\right|,
\end{equation}
Eqs.~\eqref{eq:varyO_2} and~\eqref{eq:varyP} yield
\begin{equation}
\label{eq:pdigrow_app}
\begin{aligned}
\partial_{T}^2 E_{1,O_2,z}^{(0)}
&=\kappa_{O_1,P_1}\kappa_{O_1,O_2}
\left|E_{1,O_1,z}^{(0)}\right|^2
\\
&\qquad\times E_{1,O_2,z}^{(0)}.
\end{aligned}
\end{equation}
Because the product of the coupling constants is positive, this differential equation's solution has an exponentially growing component, displayed in~\eqref{eq:inst}, and an exponentially decaying component, which will not be observed.

\section{Calculation of conversion}
\label{sec:con}
\begin{figure*}[t!]
\centering
\includegraphics[width=0.45\textwidth]{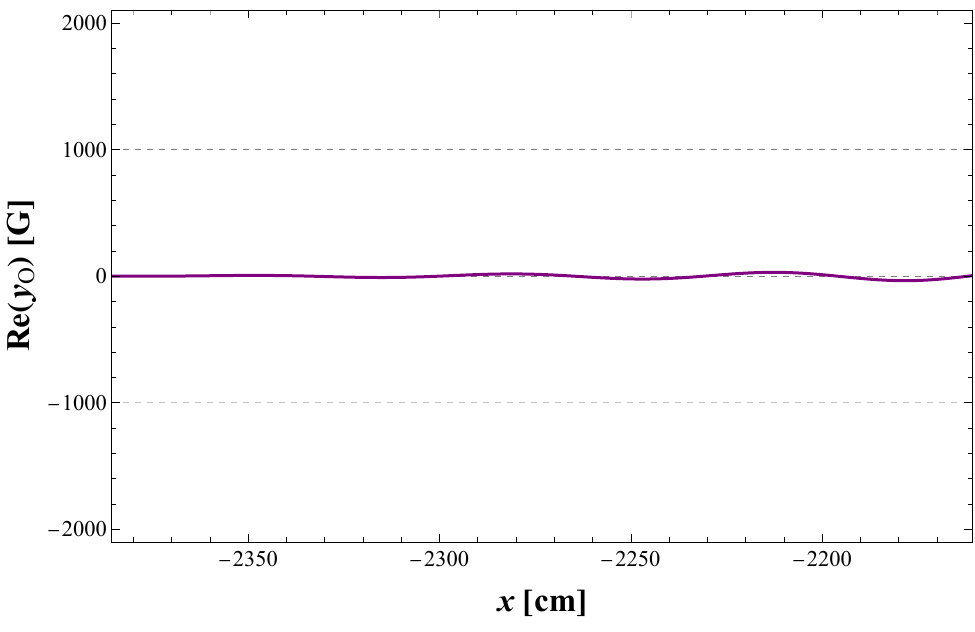}
\hfill
\includegraphics[width=0.45\textwidth]{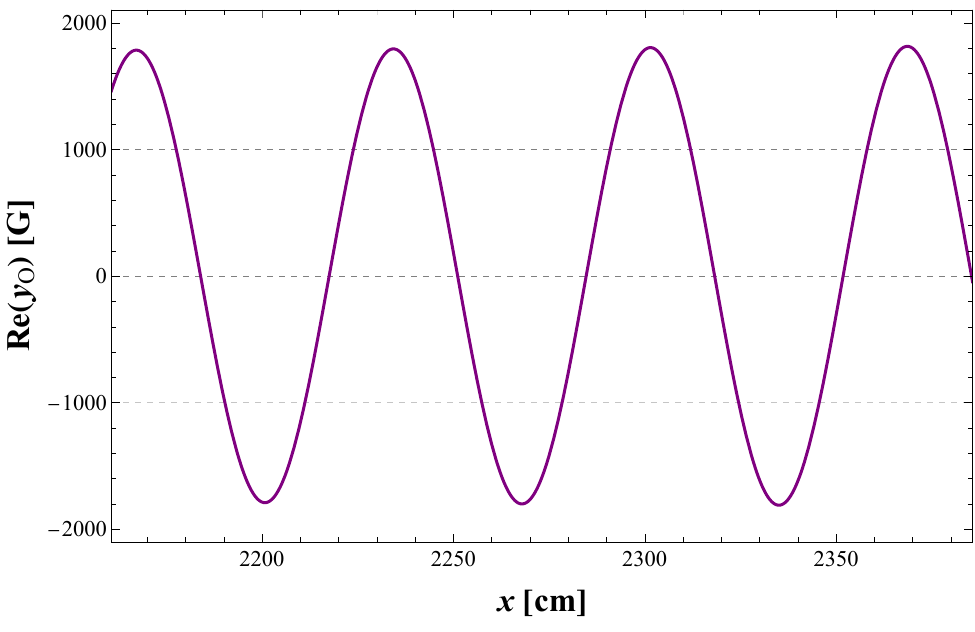}
\caption{
Real part of ordinary-like radiation before and after the conversion
surface at $x=0$ in a localized model plasma.
The plotted field amplitude includes the ordering parameter $\delta$.
Plasma parameters approximate those of the Crab pulsar magnetosphere
near the radius at which an axion with a $10\, {\rm GHz}$ Compton frequency is resonant, $r_{\rm con}=0.64 \, r_{\rm LC}$, as described further in
Appendix~\ref{sec:typ}. The plasma density is linearized about this point so that its scale length is $L_{\omega_p}
=\left|\omega_p/(\partial_x\omega_p)\right|
=6.70\times10^{7} {\rm cm}$.
The spatially constant magnetic field and initial axion density are set to
their values at $r_{\rm con}$; specific values are
$B_0=9.83\times10^{5} \,{\rm G}$ and 
$\rho_a=9.78\times10^{13} \, {\rm GeV\,cm^{-3}}$. The axion momentum and coupling constant are $\hbar k_a/(m_a c)=0.044$ and $g_{a\gamma\gamma}=10^{-12} \, {\rm GeV}^{-1}$.
}
\label{fig:conversionmodes}
\end{figure*}

The physics governing axion-photon conversion, encoded in this paper in~\eqref{eq:axionE_matrixcon}, is known in the literature \citep{raffelt1988mixing}. Here, I demonstrate this physics. I set the frequency of the axion-like field and its incoming amplitude at a boundary
\(x_L<0\), sufficiently far from the conversion surface, to values typical in a pulsar with a very significant axion cloud \citep{noordhuis2024axion}. I require that no incoming ordinary-like component be present
at \(x_L\) \citep{sommerfeld1912greensche},
\begin{equation}
\partial_x y_{O} (x_L)
=
-i k_{O0}(x_L)y_O(x_L),
\end{equation}
where
\begin{equation}
\begin{aligned}
y_O \equiv{}&
\delta E_{1,z}^{(0)}
\\
&-
\frac{\omega_{O_1}^2g_{a\gamma\gamma}B_0(x)}
{\omega_p^2(x)-\omega_{a0,rest}^2}
\delta a_1^{(0)} 
\end{aligned}
\end{equation}
is an approximate ordinary-like radiation variable. I also require that
axion-like and ordinary-like radiation flow out at a boundary \(x_R>0\),
sufficiently far from the conversion surface,
\begin{equation}
\begin{aligned}
\partial_x y_a(x_R)
&=i k_{a0} y_a(x_R), \\
\partial_x y_O(x_R)
&=i k_{O0}(x_R)y_O(x_R),
\end{aligned}
\end{equation}
where
\begin{equation}
\begin{aligned}
y_a \equiv{}&
\delta a_1^{(0)}
\\
&+
\frac{\hbar c^3 g_{a\gamma\gamma}B_0(x)}
{4\pi\left[\omega_p^2(x)-\omega_{a0,rest}^2\right]}
\delta E_{1,z}^{(0)}
\end{aligned}
\end{equation}
is an approximate axion-like radiation variable.

The solution is shown in Figure~\ref{fig:conversionmodes}. Ordinary-like
radiation is present downstream of the conversion surface even though no
incoming ordinary-like component is imposed upstream. The conversion efficiency increases
with \(B_0\) and with the density profile length scale in the manner typically understood in the literature \citep{raffelt1988mixing}. If no plasma
instability occurred, this monochromatic, linearly polarized ordinary-like radiation at frequency \(\omega_{O_1}\) would be the expected axion-photon conversion
line.

\section{Typical sizes of parameters in text and calculation of instability parameter}
\label{sec:typ}
It is helpful to calculate typical sizes of various parameters in a pulsar. Here, I present scaling estimates for various parameters in the text. I also calculate the size of the instability parameter~\eqref{eq:dampeq}, for axion conversion happening as described in Appendix~\ref{sec:con} at a radius $r_{\rm con}$ over the radial interval $r_*\leq r_{\rm con}< 1.2 r_{\rm LC}$ in a one-dimensional model of a pulsar magnetosphere.

Scaling estimates for some parameters are presented in Table~\ref{tab:pdi_scalings}. 
\begin{table*}[ht!]
\centering
\renewcommand{\arraystretch}{1.8}
\resizebox{\textwidth}{!}{%
\begin{tabular}{@{}ll@{}}
\toprule
\textbf{Quantity} & \textbf{Scaling estimate}
\tabularnewline
\midrule

$\kappa_{O_1,P_1}$ &
$\displaystyle
2.9\times10^{4} \,{\rm s^{-1}G^{-1}}
\left(\frac{\omega_{O_2}/\omega_{O_1}}{1}\right)
\left(\frac{\omega_p/\omega_{O_2}}{1}\right)^2
\left(\frac{k_{P_1}}{2\omega_{O_2}/c}\right)^{-1}
\left(\frac{m_e c_\pm^2}{170 \, {\rm keV}}\right)^{-1}
\mathcal{J}_{P}
$
\tabularnewline

$\kappa_{O_1,O_2}$ &
$\displaystyle
2.9 \, {\rm s^{-1} G^{-1}}
\left(\frac{\omega_{O_2}/\omega_{O_1}}{1}\right)
\left(\frac{k_{P_1}}{2\omega_{O_2}/c}\right)^2
\left(\frac{\sin\theta}{0.3}\right)
\left(\frac{f_i}{10^{-3}}\right)^{1/2}
\left(\frac{m_e c_\pm^2}{170 \, {\rm keV}}\right)^{1/2}
\mathcal{J}_{O}
$
\tabularnewline
$\displaystyle
\frac{
\delta^2\kappa_{O_1,P_1}\kappa_{O_1,O_2}
\left|E_{1,O_1,z}^{(0)}\right|^2
}{
\left|\partial_x\varkappa\right|
\left|v_{g,O_2}v_{g,P_1}\right|
}
$ &
$\displaystyle
0.64
\left(
\frac{\delta\left|E_{1,O_1,z}^{(0)}\right|}
{10^2 \, {\rm G}}
\right)^2
\left(
\frac{L_\varkappa}
{10^8 \, {\rm cm}}
\right)
\left(
\frac{\omega_{O_2}}
{2\pi\times10 \, {\rm GHz}}
\right)^{-1}
\left(\frac{\omega_{O_2}/\omega_{O_1}}{1}\right)^2
\left(\frac{\omega_p/\omega_{O_2}}{1}\right)^2
\left(\frac{m_e c_\pm^2}{170 \, {\rm keV}}\right)^{-1}
\mathcal{J}_{P}^{2}
$
\tabularnewline

$\displaystyle
\frac{\Delta\omega}{\omega_{O_2}}
$ &
$\displaystyle
1.7\times10^{-4}
\left(\frac{\sin\theta}{0.3}\right)
\left(\frac{f_i}{10^{-3}}\right)^{1/2}
\left(\frac{m_e c_\pm^2}{170 \, {\rm keV}}\right)^{1/2}
\left[
\frac{
7/3
}{
1+\frac{4}{3}
\left(\frac{k_{P_1}}{2\omega_{O_2}/c}\right)^2
\left(\frac{m_e c_\pm^2}{170 \, {\rm keV}}\right)
\left(\frac{\omega_p/\omega_{O_2}}{1}\right)^{-2}
}
\right]^{1/2}
\left(\frac{k_{P_1}}{2\omega_{O_2}/c}\right)
$
\tabularnewline

\bottomrule
\end{tabular}%
}
\caption{
Scaling estimates for the parametric decay of an ordinary-like wave into a
second ordinary-like wave and a phonon-like wave, presented relative to
parameters that might be found in a relativistically hot pulsar pair plasma
(note, however, that the quantities were calculated without considering
relativity).
}
\label{tab:pdi_scalings}
\end{table*}
In the table, the pair sound speed is normalized to \(c_\pm\simeq c/\sqrt{3}\),
corresponding to \(m_e c_\pm^2\simeq 170\,{\rm keV}\).
The table quantifies the fraction of the plasma composed of protons with \(f_i \equiv n_i/(n_-+n_+)\) and defines dimensionless factors \(\mathcal{J}_P\) and \(\mathcal{J}_O\),
\[
\mathcal{J}_{P}
\equiv
\frac{
1+x+\left(m_i/m_e\right)f_i
}{
1+\frac{4}{3}+\left(m_p/m_e\right)10^{-3}
}
\]
and
\[
\mathcal{J}_{O}
\equiv
\frac{
1+x+\left(m_i/m_e\right)f_i
}{
1+\frac{4}{3}+\left(m_p/m_e\right)10^{-3}
}
\left[
\frac{1+\frac{4}{3}}{1+x}
\right]^{3/2}
\]
with 
\[
\begin{aligned}
x
&\equiv
\frac{k_{P_1}^2c_\pm^2}{\omega_{p-}^2+\omega_{p+}^2}
\\
&\simeq
\frac{4}{3}
\left(\frac{k_{P_1}}{2\omega_{O_2}/c}\right)^2
\left(\frac{m_e c_\pm^2}{170\,{\rm keV}}\right)
\left(\frac{\omega_p/\omega_{O_2}}{1}\right)^{-2}.
\end{aligned}
\]
Both are equal to unity for a proton admixture \(f_i=10^{-3}\),
\(k_{P_1}=2\omega_{O_2}/c\), \(m_ec_\pm^2=170\,{\rm keV}\), and
\(\omega_p/\omega_{O_2}=1\). The table presents electric field amplitudes relative to $10^2$ G, a strength routinely excited by other plasma processes~\citep{tolman2022electric}.  The inhomogeneity scale associated with wave-number
detuning, which determines instability via~\eqref{eq:dampeq}, is
\(L_\varkappa \equiv |k_{P_1}|/\left|\partial_x \varkappa\right|\). This is normalized to the local plasma-frequency scale in Appendix~\ref{sec:con}.

I also determine if conversion processes happening in a typical strongly magnetized pulsar plasma are likely to satisfy the instability criterion identified in this paper according to the parameter~\eqref{eq:dampeq}, evaluated in Table~\ref{tab:pdi_scalings}. To do this, I take a set of profiles intended to be a one-dimensional approximation of the plasma outside a neutron star with radius $r_*$ and light cylinder radius $r_{\rm LC}$. In particular, I approximately represent the magnetic field \citep{cerutti2017electrodynamics} as
\begin{equation}
B_0\left(r \right) =
\begin{cases}
 B_* \left( \frac{r_*}{r}\right)^3,
& r \leq r_{\mathrm{LC}},
\\[10pt]
B_* \left( \frac{r_*}{r_{\rm LC}}\right)^3 \left( \frac{r_{\rm LC}}{r} \right)^2,
& r > r_{\mathrm{LC}},
\end{cases},
\label{eq:magnetic_field_profile}
\end{equation}
with $B_*$ the magnetic field at the surface of the star,
the axion density \citep{noordhuis2024axion} as
\begin{equation}
   \rho_a\left(r \right) = \rho_{a,*} \left(\frac{r_*}{r} \right)^4,
\end{equation}
with $\rho_{a,*}$ the axion density at the surface of the star,  and the plasma density as \citep{goldreich1969pulsar}
\begin{equation}
n_-\left(r \right) =
\begin{cases}
 n_* \left( \frac{r_*}{r}\right)^3,
& r \leq r_{\mathrm{LC}},
\\[10pt]
n_* \left( \frac{r_*}{r_{\rm LC}}\right)^3 \left( \frac{r_{\rm LC}}{r} \right)^2,
& r > r_{\mathrm{LC}},
\end{cases},
\label{eq:plasma_density_profile}
\end{equation}
with $n_*$ the density at the surface of the star.
(Note that although the paper does not rigorously treat varying magnetic fields and axion densities, these variations do not yield a resonant conversion region, so that they could in principle be treated with geometric optics.)

I take an axion momentum at the surface of the star that corresponds to a turning point just beyond the light cylinder according to~\eqref{eq:momrel}, 
\begin{equation}
\begin{aligned}
\frac{\hbar k_a(r_*)}{m_a c}
&=
\sqrt{\frac{2GM}{c^2}}
\\[-2pt]
&\quad\times
\left(\frac{1}{r_*}-\frac{1}{1.2r_{\rm LC}}\right)^{1/2}.
\end{aligned}
\end{equation}
(This is somewhat larger than a typical momentum~\eqref{eq:source} to allow the axions to propagate throughout the magnetosphere.)

\begin{figure*}[ht!]
    \centering
    \includegraphics[width=0.5\textwidth]{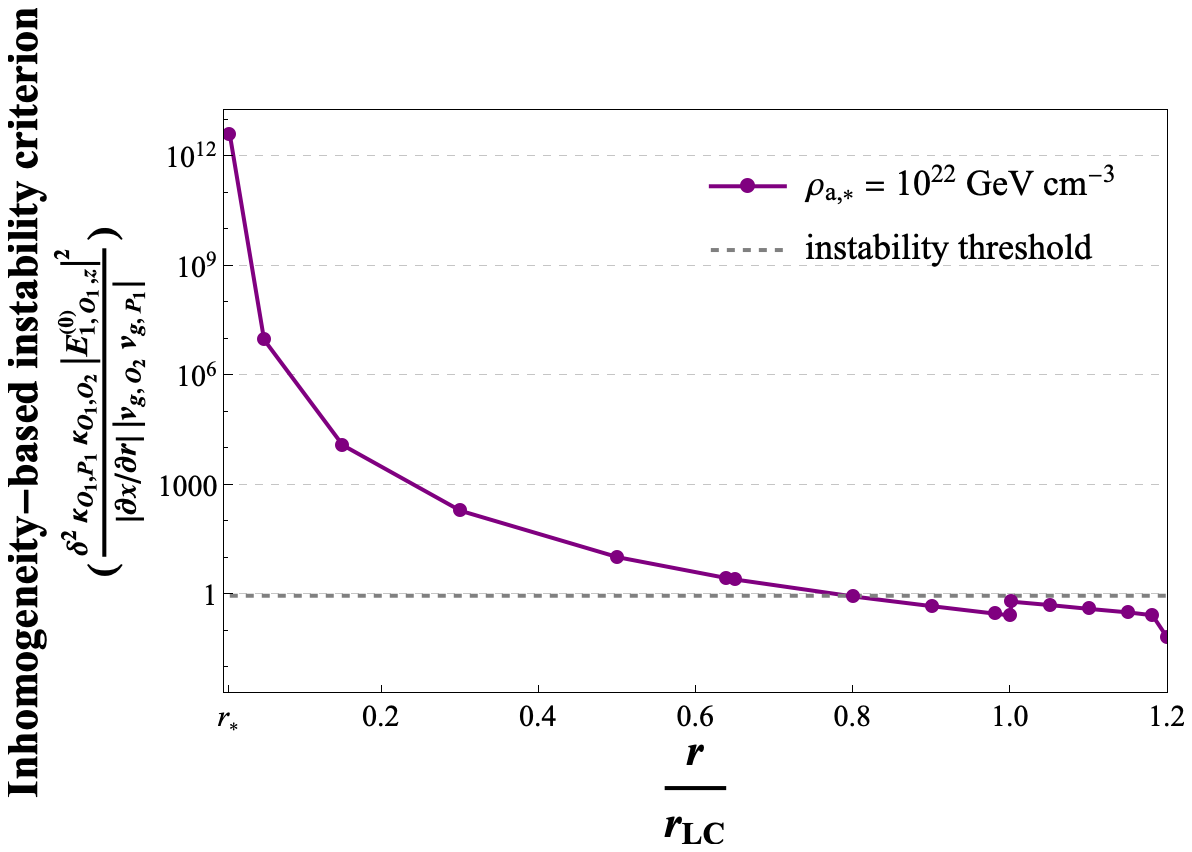}
\caption{
Instability criterion~\eqref{eq:dampeq} calculated for
$\rho_{a,*}=10^{22}\,{\rm GeV\,cm^{-3}}$ over
$r_*\leq r_{\rm con} < 1.2\,r_{\rm LC}$.
The adopted Crab-like pulsar parameters are
$r_*=10\,{\rm km}$, $M=1.4\,M_\odot$,
$P = 0.033 \, {\rm s}$, $B_*=10^{12}\,{\rm G}$, and
$n_{*}=6.31\times10^{17}\,{\rm cm^{-3}}$.
At each radius, the axion mass is chosen so that
$m_a c^2/\hbar=\omega_p(r_{\rm con})$, and the ordinary-like amplitude
produced by the local conversion calculation is used in the instability
criterion.
}

    \label{fig:pdiparam}
\end{figure*}

Over the radial interval $r_*\leq r_{\rm con}< 1.2 r_{\rm LC}$, I calculate the conversion of an axion population that would be resonant at each location, i.e., one with $m_a c^2/\hbar =  \omega_{p}\left(r_{\rm con}\right)$, by running the calculation described in Appendix~\ref{sec:con}, where $x=r-r_{\rm con}$ is the distance from the resonant point. Then, I use the amplitude of the ordinary-like radiation that results and the plasma profiles to evaluate the PDI instability criterion~\eqref{eq:dampeq}, which is valid locally as described in \cite{rosenbluth1972parametric}.  The results are shown in Figure~\ref{fig:pdiparam} for an axion cloud size estimated in recent literature \citep{noordhuis2024axion}. 

Table~\ref{tab:pdi_scalings} shows that the fractional spacing of the instability-driven
bands, \(\Delta\omega/\omega_{O_2}\), is typically small. Figure~\ref{fig:pdiparam} shows that the parametric instability should be easy to excite at plausible astrophysical parameters. These estimated quantities cannot, however, be compared exactly with astrophysical observations because I neglect relativistic and kinetic effects.
\nocite{Landau1932,Zener1932,rubbmark1981dynamical}
\bibliographystyle{apsrev4-2}
\bibliography{bib}

\end{document}